\documentstyle[prl,aps,preprint,epsfig]{revtex}
\begin{document}
\tighten

\title{Simplification of Higher-Twist Evolution in \\
the Large $N_c$ Limit: Why and Why Not}
\author{Xiangdong Ji and Jonathan Osborne}
\bigskip

\address{
Department of Physics \\
University of Maryland \\
College Park, Maryland 20742 \\
{~}}

\date{UMD PP\#99-045 ~~~DOE/ER/40762-1169~~~ November 1998}

\maketitle

\begin{abstract}
Working in the light-cone gauge, we find a simple 
procedure to calculate the autonomous one-loop
$Q^2$ evolution of the twist-three part of 
the nucleon $g_T(x, Q^2)$ structure function
in the large-$N_c$ limit. Our 
approach allows us to investigate the 
possibility of a similar large-$N_c$ 
simplification for other higher-twist 
evolutions. In particular,
we show that it does not occur 
for the twist-four part of the
$f_4(x, Q^2)$, $ g_3(x, Q^2)$ and $h_3(x, Q^2)$ distributions. 
We also argue that the simplification of the 
twist-three evolution
does not persist beyond one loop.

\end{abstract}
\pacs{xxxxxx}

\narrowtext
Feynman's parton model of incoherent parton
scattering provides a transparent 
picture  of what happens in a broad class of
high-energy scattering processes. Modulo 
field theoretical logarithms, the parton 
model can be derived in quantum chromodynamics (QCD)
in the form of factorization theorems \cite{collins}. 
Better yet, QCD allows us to go beyond 
the naive parton model by consistently including
the effects of the parton transverse 
momentum and coherent parton scattering.  
A simple example of coherent 
parton scattering is the interference
of a single quark with a
quark {\it and} a gluon in a nucleon target. To describe this 
phenomenon, it is necessary to introduce
a three-parton light-cone correlation function
\begin{equation}
 M^\alpha (x,y,Q^2) = \int {d\lambda\over 2\pi} {d\mu\over 2\pi}
     e^{i\lambda x} e^{i\mu(y-x)}
   \langle PS| \bar \psi(0)iD^\alpha(\mu n)\psi(\lambda n)
     |PS\rangle \ ,  
\end{equation}
where $n$ is a light-cone vector, $\psi$ a
quark field, and $|PS\rangle$ the nucleon state.
The general parton correlations involve
more than one Feynman variable, and hence
their scale ($Q^2$) evolution 
is more complicated than the usual 
Dokshitzer-Gribov-Lipatov-Altarelli-Parisi (DGLAP) 
evolution equations for the Feynman 
parton densities. Technically, the complication
arises from the so-called higher-twist part 
of the correlations. 
  
Experimental study of parton correlations is 
challenging for a number of reasons. One  
is the lack of processes in which all Feynman variables in
a parton correlation can be kinematically controlled. 
For instance, in polarized lepton-nucleon deep-inelastic 
scattering (DIS), one can measure the 
structure function $g_T(x,Q^2)$. In the Bjorken limit, $g_T(x, Q^2)$ 
is related to a $y$-moment of the above 
correlation function. Since a moment of $M_\alpha(x,y, Q^2)$
does not evolve autonomously, knowing the entire
$g_T(x, Q^2)$ at one scale is not 
sufficient to calculate it at another. This makes
an analysis of $g_T(x,Q^2)$ data at 
different scales difficult. 

Several years ago, Ali, Braun, and Hiller (ABH) \cite{ali}
made a remarkable discovery that in 
the limit of the large number of color $N_c$, 
the twist-three part of $g_T(x, Q^2)$ does evolve autonomously 
at the one-loop level. The result has since been widely
used in model calculations and analyses of 
experimental data \cite{use}. More recently similar
results have been found for the evolutions of other twist-three
functions $h_L(x, Q^2)$ and $e(x, Q^2)$\cite{other}. 
Given the practical importance of the ABH result, a deeper 
understanding of the large $N_c$ simplification is 
clearly desirable. Moreover, it is interesting
to investigate the possibility of a similar 
simplification 
at two or more loops and for analogous
twist-four correlations. 

In this paper we calculate directly the large-$N_c$
evolution of $g_T(x, Q^2)$ in the light-cone gauge. 
We find that the autonomy of the twist-three evolution
arises from a special property of one particular 
Feynman diagram. Since this property is independent of 
the $\gamma$-matrix structure of the composite operators inserted, 
the ABH result generalizes immediately to the 
twist-three parts of $h_L(x, Q^2)$ and $e(x, Q^2)$.
Unfortunately, for various reasons we shall explain, 
there is no similar large-$N_c$ simplification 
for twist-four functions, nor for $g_2(x, Q^2)$ 
beyond one loop. 

We begin our discussion with a brief introduction to 
the $g_T(x,Q^2)$ structure function of the nucleon. 
In inclusive DIS, all information about the nucleon structure 
is summarized in the following hadron tensor,
\begin{equation}
   W^{\mu\nu}(P,S,q) = {1\over 4\pi}\int d^{\,4} \xi\, e^{iq\cdot \xi} \langle PS|
   [J_\mu(\xi), J_\nu(0)]|PS \rangle \ ,  
\end{equation}
where $J^\mu =\sum_q e_q^2 \bar \psi_q\gamma^\mu \psi_q$ is 
the electromagnetic current and $q$ is the spacelike
virtual photon momentum.  The antisymmetric part of the
hadron tensor, $W^{[\mu\nu]}$, is polarization-dependent 
and can be characterized in terms of  the
two structure functions $g_1(x_B,Q^2)$ and $g_2(x_B,Q^2)$:
\begin{equation}
 W^{[\mu\nu]}  = -i\epsilon^{\mu\nu\alpha\beta}
        q_\alpha \left(S_\beta {g_1(x_B,Q^2)\over \nu}
     + [\nu S_\beta-(S\cdot q)P_\beta]{g_2(x_B,Q^2)\over \nu^2}\right) \ , 
\end{equation}
where we have chosen the kinematic factors so that
$g_1(x_B,Q^2)$ and $g_2(x_B,Q^2)$ survive the scaling limit 
$Q^2=-q^2\rightarrow \infty$, $\nu=P\cdot q\rightarrow\infty$
and $x_B = Q^2/2\nu$ = finite. 
In Feynman's parton model,
$g_1(x_B, Q^2)$ is related to the parton 
helicity density $\Delta q_a(x, Q^2)$
\begin{equation}
    g_1(x_B,Q^2) = {1\over 2} \sum_a e_a^2 
   \left[\Delta q_a(x_B, Q^2) + \Delta q_a(-x_B,Q^2) \right] \ , 
\end{equation}  
where $e_a$ is the electric charge and $a$ sums over 
light quark species. 

The structure function $g_2(x_B, Q^2)$, however, does not have a 
simple parton model interpretation. Defining
$g_T(x_B, Q^2) = g_1(x_B, Q^2) + g_2(x_B, Q^2)$, 
an operator-product-expansion analysis yields 
\cite{ope} 
\begin{equation}
    g_T(x_B,Q^2) = {1\over 2} \sum_a e_a^2 
   \left(\Delta q_{Ta}(x_B, Q^2) + \Delta q_{Ta}(-x_B, Q^2)\right) \ , 
\end{equation} 
where we have neglected all power and radiative corrections and 
\begin{equation}
  \Delta q_{Ta}(x, Q^2) = {1\over 2M} 
  \int {d\lambda \over 2\pi} 
   e^{i\lambda x} \langle PS_\perp|\bar 
   \psi_a(0) \gamma^\perp\gamma_5 \psi_a(\lambda n)
    |PS_\perp \rangle \ . 
\end{equation}
The trouble with a parton model interpretation of 
$\Delta q_{Ta}(x, Q^2)$ 
can easily be seen in light-front
quantization in which only the ``good''
component of the Dirac field $\psi_+ = P_+\psi$ 
has a simple Fock expansion ($P_\pm = 
\gamma^\mp \gamma^\pm/2$, $\gamma^\pm = 
(\gamma^0\pm\gamma^3)/\sqrt{2}$), whereas the ``bad''
component $\psi_- = P_-\psi$ is 
constrained by the following equation of motion
\begin{equation}
   \psi_-(\lambda n) =- {1\over 2} {1\over in\cdot \partial}
     \not\! n i\!\not\!\! D_\perp (\lambda n)\psi_+(\lambda n) \ . 
\label{eom}
\end{equation}
[In some sense $\psi_-$ represents a quark-gluon composite.] 
Unlike $\Delta q_a(x, Q^2)$, $\Delta q_{Ta}(x, Q^2)$ contains 
a bad component because of the $\gamma^\perp$. 

For the same reason, the scale evolution 
of $\Delta q_{Ta}(x, Q^2)$ is now more intricate than that 
of $\Delta q_a(x, Q^2)$. Its $n$-th moment is written
\begin{equation}
   \int^1_{-1} \Delta q_{Ta}(x, Q^2) x^{n} dx 
   = {1\over 2 M}n_{\mu_1}\cdots n_{\mu_n}
    \langle PS_\perp|\theta^{\perp(\mu_1\cdots\mu_{n})}
     |PS_\perp\rangle,
\end{equation}
where $\theta^{\sigma(\mu_1\cdots\mu_{n})}
= \bar \psi\gamma^\sigma iD^{(\mu_1}\cdots iD^{\mu_{n})}\psi$,
with $(\mu_1\cdots\mu_{n})$ indicating symmetrization 
of the indices and removal of the traces. 
The $\theta$-operator contains both 
twist-two $\theta^{(\sigma\mu_1\cdots\mu_{n})}$
(totally symmetric and traceless) and twist-three 
$\theta^{[\sigma(\mu_1]\mu_2\cdots\mu_{n})}$ (mixed symmetric
and traceless) contributions, where $[\sigma \mu_1]$ 
denotes antisymmetrization. 
It turns out, however, that for a given symmetry structure there
are multiple twist-three operators. In fact, a complete 
basis of these operators was first identified in \cite{shuryak},
\begin{eqnarray}
&& R_i^n = \bar \psi iD^{(\mu_1} 
  \cdots iD^{\mu_{i-1}} (-ig)F^{\sigma \mu_i} iD^{\mu_{i+1}}
  \cdots iD^{\mu_{n-1}} \gamma^{\mu_{n})}\gamma_5\psi
  \nonumber \\
&& S_i^n = \bar \psi iD^{(\mu_1} 
  \cdots iD^{\mu_{i-1}} g\tilde F^{\sigma \mu_i} iD^{\mu_{i+1}}
  \cdots iD^{\mu_{n-1}} \gamma^{\mu_{n})}\psi\ , 
\end{eqnarray}
where $i=1, ..., n-1$. 
The operator $\theta^{[\sigma(\mu_1]\mu_2\cdots\mu_{n})}$
is just a special linear combination of them,
\begin{equation}
  \theta^{[\sigma(\mu_1]\mu_2\cdots\mu_{n})} ={1\over 2(n+1)} \sum_{i=1}^{n-1} 
   (n-i)(R_i^n -R_{n-i}^n+S_i^n+S_{n-i}^n) \ . 
\label{relation}
\end{equation}
The anomalous dimension matrix in the above operator basis was
first worked out by Bukhvostov et al. and 
later reproduced by a number of authors with different
methods\cite{matrix}. The result is what one would generally
expect.  To evolve the matrix element of 
$\theta^{[\sigma(\mu_1]\mu_2\cdots\mu_n)}$, it is 
not enough just to know it at an initial scale---one must know 
all the matrix elements of $W_i^n=R_i^n-R_{n-i}^n+S_i^n+S_{n-i}^n$ there. 

By studying the anomalous dimension matrix 
in the large $N_c$ limit, Ali, Braun and Hiller found that
the eigenvector corresponding to the lowest
eigenvalue is just the linear combination
of twist-three operators on the right-hand side
of Eq. (\ref{relation}). In other words,  
the twist-three part of $\Delta q_{Ta}(x, Q^2)$ evolves 
autonomously in this limit.
To better understand ABH's result, we 
calculate the large-$N_c$ evolution of 
$\Delta q_{Ta}(x, Q^2)$ directly. We start
with the mixed-twist operator
$\theta^{\sigma(\mu_1\mu_2\cdots\mu_{n})}$ in Eq. (8)
and look for possible divergences when 
inserted in multi-point Green's functions. 
To reduce the number of Feynman diagrams, we choose the 
light-cone gauge $A^+=0$ and take the $\perp +\cdots +$
component of the $\theta$-operator. Let's call the resulting
operator $\theta_n \equiv \bar \psi \gamma^\perp\gamma_5 
(i\partial^+)^{n}\psi$, and its twist-two and twist-three
parts $\theta_{n2}$ and $\theta_{n3}$, respectively. The 
Feyman rule for $\theta_n$ is simply $\gamma^\perp\gamma^5
(k^+)^{n}$, where $k$ is the momentum of the quark. 

By light-cone power counting, we need only 
consider two- and three-point functions. 
Since the external lines carry color, 
we must ask what type of diagrams 
dominates the large $N_c$ limit. The simple rule 
we find is that when all external lines 
are drawn to one point (infinity), 
the planer diagrams are leading.
All one-particle-irreducible
(1PI) leading diagrams with one 
$\theta$ insertion are shown in Fig. 1. 

The ultraviolet divergences in the two point 
Green's function can obviously be subtracted 
with the matrix element of $\theta_n$ itself. 
The only diagram in which the divergences may 
not be subtracted by $\theta_n$
is Fig. 1b. An explicit calculation 
shows that the ultraviolet divergences 
correspond to the following local operator:
\begin{eqnarray}
  && {1\over 2}C_A {g^2\over 8\pi^2}\ln Q^2
   \left[ -{1\over (n+2)} \sum_{i=0}^{n-1}
    \bar \psi \not\! n \gamma_5 (i\partial^+)^i iD^\perp
        (i\partial^+)^{n-1-i} \psi \right.\nonumber 
  \\
    + && \left.\left(\sum_{i=1}^{n+1}{1\over i}-{1\over 2(n+1)}\right)
     \left(\bar \psi i\!\!\not\!\!D_\perp\not\! 
   n \gamma^\perp\gamma_5(i\partial^+)^{n-1}
   \psi  + \bar \psi 
   (i\partial^+)^{n-1}\gamma^\perp\gamma_5 \not\! n i\!\!\not\!\! D_\perp 
   \psi\right)\right] \ , 
\label{result}
\end{eqnarray}
where we have neglected the contributions of light-cone 
singularities which will be cancelled eventually. 
Notice that the first term is present in the  
twist-two operator 
\begin{equation}
    \theta_{n2}  = {1\over n+1}\left(
    \bar \psi \gamma^\perp \gamma_5(i\partial^+)^{n}\psi
     + \sum_{i=0}^{n-1}
    \bar \psi \gamma^+\gamma_5 (i\partial^+)^iiD^\perp
      (i\partial^+)^{n-i-1}\psi\right) \ ,
\end{equation}
\begin{figure}
\label{fig1}
\epsfig{figure=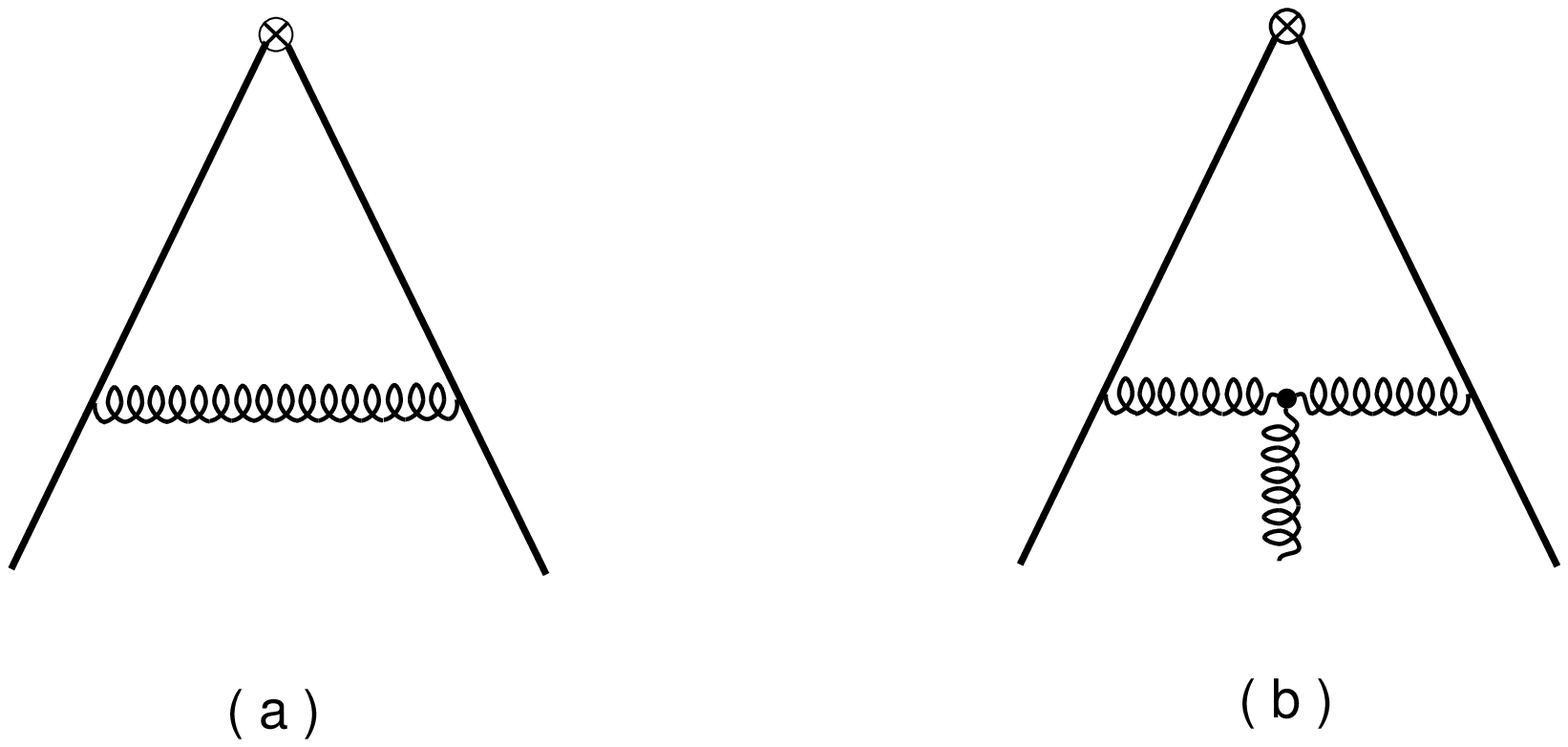,height=4cm}
\caption{Two and three-point 1PI Feynman diagrams contributing
to the evolution of $\theta_n$ in the large $N_c$ limit.}
\end{figure}    
\noindent
and the remaining two terms
can be converted to $\theta_n$ by using the 
equation of motion in Eq. (\ref{eom}). Thus
we easily arrive at the ABH conclusion that $\theta_{n3}$ 
evolves autonomously in the large-$N_c$ limit. 

Including the contribution from Fig. 1a as well as
the one-particle-reducible ones that cannot be neglected
in the light-cone gauge, we obtain the following equation,
\begin{equation}
  { d\theta_n\over d\ln Q^2}
  = {\alpha_s(Q^2) \over 2\pi}{ N_c\over 2}
\left[{n+1\over n+2}\theta_{n2} + 
   \left (-2\sum_{i=1}^{n+1}{1\over i} + {1\over n+1} + {1\over 2}\right)
    \theta_n\right] \ . 
\end{equation}
Separating out the twist-two and twist-three parts, 
we not only recover the well-known twist-two evolution, 
but also the twist-three result
\begin{equation}
    { d\theta_{n3}\over \ln Q^2}
   ={ \alpha_s(Q^2)\over 2\pi}\left(-2\sum_{i=1}^{n+1}
   {1\over i}+ {1\over n+1} + {1\over 2}\right)\theta_{n3},
\end{equation}
which is identical to the result in Ref. \cite{ali}. 

It is quite clear that the $i$-independence of 
the coefficients in the sum of Eq. (\ref{result})
is the key for the autonomous evolution of $\theta_{n3}$. 
On the other hand, this property is not totally unexpected 
if one inspects Fig. 1b more closely. Interpreting 
this diagram in the coordinate space, we see that the internal
gluon propagates {\it homogeneously} from one quark to the other. 
By homogeneously, we mean that at
any point along the path of the propagation,
the gluon behaves exactly the same way, 
except, of course, at the points where the gluon 
and quarks interact. Now the spatial location 
of the interaction with the external gluon 
determines the number of derivatives before 
and after the gluon field in the subtraction 
operator. Since the internal gluon propagation is 
homogenous, different locations of the triple-gluon 
vertex should produce similar physical effects.  
Therefore, the coefficients of the
subtraction operators $\bar \psi \not\! n 
\gamma_5 (i\partial^+)^i iD^\perp (i\partial^+)^{n-1-i}
\psi$ should be independent of $i$. On the other hand,
the two extra terms in Eq.(\ref{result})
correspond to the triple-gluon vertex just next to 
the external quark lines, where the homogeneity 
is lost. 

Since the homogeneous property  of the internal
gluon line is independent of the gamma matrix structure 
of the operator inserted, we conclude that the other 
twist-three distributions
$e(x, Q^2)$ and $h_L(x, Q^2)$ evolve also autonomously in the
large $N_c$ limit. A quick calculation confirms
the evolution equations found in Ref. \cite{other}.

Encouraged by the success of the above approach, 
we apply it to the analogous twist-four evolution.
In Ref. \cite{jaffe}, the three one-variable
distributions $f_4(x, Q^2)$, $g_3(x, Q^2)$ and $h_3(x, Q^2)$ 
are shown to contain twist-four. 
For example, $f_4(x)$ is defined as
\begin{equation}
      f_4(x) = {1\over M^2}
   \int {d\lambda \over 2\pi} \langle P |\bar \psi(0)
    \gamma^- \psi(\lambda n) |P\rangle \ .
\end{equation}
It was shown in Ref. \cite{ji} that $f_4(x, Q^2)$ 
contributes to the $1/Q^2$ term of the longitudinal 
scaling function $F_L$  of the nucleon
\begin{equation}
    F_L(x_B, Q^2) ={ 2x^2_B M^2\over Q^2}  \sum_a 
       e_a^2 f_{4a}(x_B, Q^2)\ ,  
\end{equation}
where we have neglected higher-order 
radiative corrections. Here, autonomous
evolution of $f_4(x, Q^2)$ would  
simplify the analysis of 
$F_L$ data immensely. 

\begin{figure}
\label{fig2}
\epsfig{figure=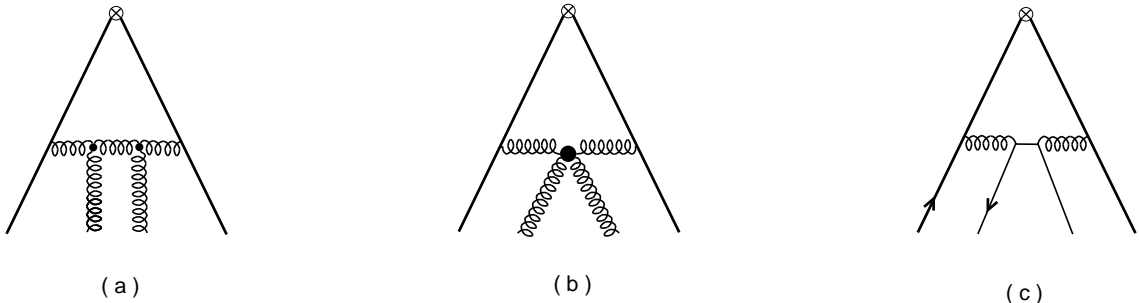,height=4cm}
\caption{Four-point 1PI Feynman diagrams contributing
to the evolution of $\hat O$ in the large $N_c$ limit.}
\end{figure}    
 In the large $N_c$ limit, we consider one insertion
of the operator $\hat O =\bar \psi\gamma^-(i\partial^+)^{n}\psi$ 
into two-, three- and four-point Green's functions. At one-loop 
order, the 1PI two- and three-point diagrams 
are identical to those in Fig. 1 and the 1PI 
four-point diagrams are shown in Fig. 2. Only the 
three and four point diagrams can potentially 
destroy the autonomous evolution of $\hat O$. 
Let us start with Fig. 2a. One of the divergent 
contributions from this diagram introduces
the following local subtraction
\begin{equation}
  \sum_i \bar \psi i\!\not\!\! D_\perp \not \! n 
  (i\partial^+)^i i\!\not\!\! D_\perp (i\partial^+)^{n-i-2}\psi
   + {\rm h. c.}
\end{equation}
where all the coefficients are independent of $i$
again because of the homogeneity of the gluon 
propagator. Using the equation of motion, we can 
write this as
\begin{equation}
    \sum_i \bar \psi 
         (i\partial^+)^i i\not\!\! D_\perp (i\partial^+)^{n-i-2}\psi
   + {\rm h. c.}
\end{equation}
Since this operator cannot be reduced to either the twist-two or 
twist-four part of $\hat O$, the evolution
of the latter cannot be autonomous unless this contribution
is cancelled by other diagrams. The only other diagram 
containing the same divergence structure is Fig. 1b 
with an insertion of $\hat O$. 
Unfortunately, our explicit calculation did not 
produce this cancellation. The same phenomenon
occurs for the twist-four
part of $g_3(x, Q^2)$ and $h_3(x, Q^2)$. 

Thus, the large $N_c$ simplification seems to happen
only for the evolution of the 
twist-three part of $g_T(x,Q^2)$, $h_L(x,Q^2)$ 
and $e(x, Q^2)$.  Does it happen for them 
at two and higher loops?
In Fig. 3, we show two examples of Feynman 
diagrams that we suspect break 
the autonomy of the $\theta_{3n}$-evolution, i.e., 
they may contain divergences that cannot be
subtracted by $\theta_{n2}$ and $\theta_{n3}$ only. 
Our suspicion is based on the inhomogeneity of
the gluon progator. The internal gluon that 
propagates from one quark to another has different 
wavelengths in the different parts of the 
propagation. Its interaction with the 
external gluon is different at different 
spatial locations. Thus the subtraction operators
have different coefficients depending on 
the number of derivatives before and
after the external gluon field. An explicit
calculation of Fig. 3a confirms our suspicion.

\begin{figure}
\label{fig3}
\epsfig{figure=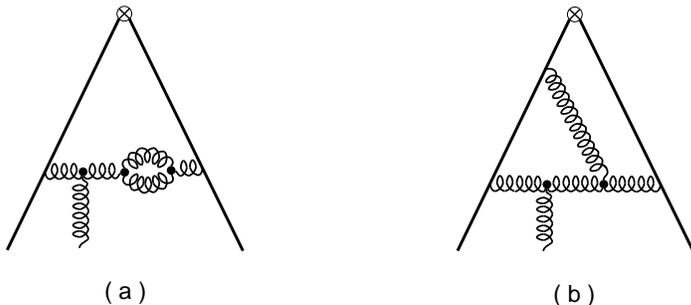,height=4cm}
\caption{Some two-loop 1PI Feynman diagrams that might break
the autonomy of the $\theta_n$ in the large $N_c$ limit.}
\end{figure}    
This leaves us with only one possibility for autonomous  
two-loop evolution of $\theta_{n3}$: the unwanted structures
cancel in the sum of all large-$N_c$ two-loop 
diagrams.
Calculating all those diagrams is a big 
task. However, even without
an explicit calculation, we do not expect 
the cancellation to happen.
The fundamental reason is that large $N_c$
represents only a selection of a subset
of Feynman diagrams, whereas the result 
of an individual diagram is independent 
of the large-$N_c$ limit. Cancellations 
of a structure do not happen among 
Feynman diagrams unless there is a 
symmetry.

Therefore we conclude that the autonomy 
of one-loop evolution for a set of special 
twist-three distributions at large $N_c$ seems 
accidental. In the light-cone gauge, it 
can be easily traced to a special property of Fig. 1b. 
The simplification does not happen for 
the analogous twist-four distributions at one loop, 
nor for those twist-three distributions 
at two or higher loops. Nonetheless, the discovery
of Ali, Braun, and Hiller remains as a significant
step forward in the study of the $g_2(x, Q^2)$
structure function. Without the autonomous
one-loop evolution, an analysis of experimental data
on the twist-three contribution
would be severely constrained. 
 
\acknowledgements
J.O. acknowledges a useful conversation on the subject of
the paper with A. Belitsky. This work is supported in 
part by funds provided by the
U.S.  Department of Energy (D.O.E.) under cooperative agreement
DOE-FG02-93ER-40762.

\end{document}